\documentclass{aa}  
\usepackage{graphicx}
\usepackage{txfonts}
\usepackage[figuresright]{rotating}
\usepackage{natbib}
\bibpunct{(}{)}{;}{a}{}{,}
\usepackage{graphicx}   
\usepackage{verbatim}   
\usepackage{color}      
\begin{document}

   \title{Polycyclic aromatic hydrocarbon processing in a hot gas}

   \author{E. R. Micelotta\inst{1,2}, A. P. Jones\inst{2}, A. G. G. M. Tielens\inst{1,3}}

   \offprints{E. R. Micelotta}

   \institute{Sterrewacht Leiden, Leiden University, P.O. Box 9513, 2300 RA 
              Leiden, The Netherlands\\
              \email{micelot@strw.leidenuniv.nl}
              \and
              Institut d'Astrophysique Spatiale, Universit\'{e} Paris Sud and CNRS (UMR 8617),
              91405 Orsay, France
              \and
              NASA Ames Research Center, MS 245-3, Moffett Field, CA 94035, USA\\ 
             }

   \date{Received 20 January 2009; accepted 16 October 2009}

  \abstract
   {PAHs are thought to be a ubiquitous and important dust component of the interstellar medium. 
    However, the effects of their immersion in a hot (post-shock) gas have never before been fully
    investigated.}
   {We study the effects of energetic ion and electron collisions on PAHs in the hot
    post-shock gas behind interstellar shock waves.}
   {We calculate the ion-PAH and electron-PAH nuclear and electronic interactions, above the
    carbon atom loss threshold, in {{\sc H\,ii}} regions and in the hot post-shock gas for temperatures
    ranging from $10^{3}-10^{8}$ K.}
   {PAH destruction is dominated by He collisions at low temperatures ($T <$ 3$\times$10$^4$ K), 
     and by electron collisions at higher temperatures. Smaller PAHs are destroyed faster for
     $T <$ 10$^6$ K, but the destruction rates are roughly the same for all PAHs at higher
     temperatures. The PAH lifetime in a tenuous hot gas ($n_{\rm H} \approx 0.01$ cm$^{-3}$, 
     $T \approx$ 10$^7$ K), typical of the coronal gas in galactic outflows, is found to be 
     about thousand years, orders of magnitude shorter than the typical lifetime of such
     objects.}
   {In a hot gas, PAHs are principally destroyed by electron collisions and not by the 
     absorption of X-ray photons from the hot gas. The resulting erosion of PAHs occurs 
     via C$_2$ loss from the periphery of the molecule, thus preserving the aromatic
     structure. The observation of PAH emission from a million degree, or more, gas is 
     only possible if the emitting PAHs are ablated from dense, entrained clumps that 
     have not yet been exposed to the full effect of the hot gas.}

   \keywords{shock waves -- dust, extinction  -- ISM: jets and outflows}
   \authorrunning{E. R. Micelotta et al.}

   \titlerunning{PAH processing in a hot gas}
   \maketitle


\section{Introduction}

The mid-infrared spectral energy distribution of the general interstellar 
medium of galaxies is dominated by strong and broad emission features at 
3.3, 6.2, 7.7 and 11.3 $\mu$m. These features are now univocally attributed 
to vibrational fluorescence of UV pumped, large ($\simeq 50$ C-atoms) 
Polycyclic Aromatic Hydrocarbon (PAHs) molecules. These large molecules 
are very abundant ($3\times 10^{-7}$ by number relative to H-nuclei) and ubiquitous 
in the ISM \citep[for a recent review see][]{tielens08}. Besides large PAH 
molecules, the spectra also reveal evidence for clusters of PAHs -- containing 
some hundreds of carbon atoms -- and very small grains ($\simeq 30$ \AA ). 
Indeed, PAHs seem to represent the extension of the interstellar dust size 
distribution into the molecular domain \citep[e.g.][]{desert90, draine01}.

PAH molecules are an important component of the ISM, for example, dominating 
the photoelectric heating of neutral atomic gas and the ionization balance of 
molecular clouds. Small dust grains and PAHs can also be important agents in 
cooling a hot gas, at temperatures above $\sim 10^6$ K \citep[e.g.][]{dwek87}, 
through their interactions with thermal electrons and ions. The energy transferred 
in electron and ion collisions with the dust is radiated as infrared photons. 
The evolution of dust in such hot gas ($T \ga 10^6$ K), e.g., within supernova 
remnants and galactic outflows, is critical in determining the dust emission 
from these regions and therefore the cooling of the hot gas. The destruction 
of PAHs and small dust grains in a hot gas may also be an important process 
in the lifecycle of such species \citep{dwek96, jones96}.

Observationally, there is little direct evidence for PAH emission
unequivocally connected to the hot gas in supernova
remnants. \citet{reach06} have identified four supernova remnants with
IR colors that may indicate PAH emission.  \citet{tappe06} have
detected spectral structure in the emission characteristics of the
supernova remnant N132D in the Large Magellanic Cloud that they
attribute to spectral features of PAHs with sizes of $\simeq 4000$
C-atoms. Bright 8 $\mu$m emission has been observed by IRAC/Spitzer
associated with the X-ray emission from the stellar winds of the
ionizing stars in the M 17 {\sc H\,ii } region \citep{povich07}. Likely, this
emission is due to PAHs -- probably, in entrained gas ablated from the
molecular clouds to the North and West of the stellar
cluster. Finally, bright PAH emission has been detected associated
with the hot gas of the galactic wind driven by the starburst in the
nucleus of the nearby irregular galaxy, M 82 \citep{engel06, beirao08,
galliano08}.

Electron and ion interactions with dust and the implications of those interactions 
for the dust evolution and emission have already been discussed in the literature
\citep[e.g.][]{draine79, dwek87, jones94, dwek96, jones96}. 
In this work we extend this earlier work to the case for PAHs, using our 
study of PAH evolution due to ion and electron interactions in shock waves in the ISM 
\citep[][ hereafter MJT]{micelotta09}. Here we consider the fate of PAHs in the hot gas 
behind fast non-radiative shocks and in a hot gas in general.

The aim of this paper is to study the PAH stability against electron and ion 
collisions (H, He and C) in a thermal gas with temperature $T$ in the range 
10$^{3}$~--~10$^{8}$ K. 

The paper is organized as follows:
Sect. 2 and Sect. 3 describe the treatment of ion and electron interactions with PAHs, 
Sect. 4 illustrates the application to PAH processing in a hot gas and Sect. 5 presents 
our results on PAH destruction and lifetime. The astrophysical implications are 
discussed in Sect. 6 and our conclusions summarized in Sect. 7.

\section{Ion interaction with PAHs}

\subsection{Electronic interaction}

The ion -- PAH collision can be described in terms of two simultaneous 
processes which can be treated separately \citep{lind63}: \textit{nuclear 
stopping} or \textit{elastic energy loss} and \textit{electronic stopping} 
or \textit{inelastic energy loss}.

The nuclear stopping consists of a binary collision between the incoming
ion (projectile) and a single atom in the target material. A certain amount 
of energy will be transferred directly to the target atom, which will be 
ejected if the energy transferred is sufficient to overcome the threshold 
for atom removal. The physics of the nuclear interaction for a PAH target 
was presented in the companion paper (MJT). A summary of 
the theory is provided in Sect. 2.2 and we here present the results of our 
calculation.

In this paper we focuse on the electronic stopping, which consists of the 
interaction between the projectile and the electrons of the target PAH, with a 
subsequent energy transfer to the whole molecule. The resulting electronic 
excitation energy will be transferred to the molecular vibrations of the PAH 
through radiationless processes (eg., interconversion \&\ intramolecular 
vibrational redistribution). The vibrationally excited molecule will decay 
through either (IR) photon emission or through fragmentation (i.e., H-atom 
or C$_2$H$_{n}$ loss, where $n$ = 0, 1, 2).

No specific theory describes the energy transfer to a PAH via electronic
excitation, so we adopt the same approach developed by e.g. \citet{sch99}
and \citet{hadjar01} who modelled electronic interactions in fullerene, C$_{60}$. 
To calculate the energy transferred to a PAH, we treat the large number of 
delocalized valence electrons in the molecule as an electron gas, where the 
inelastic energy loss of traveling ions is due to long range coupling to 
electron-hole pairs \citep{ferr79}. 

In the energy range we consider for this study, the energy transferred scales 
linearly with the velocity $v$ of the incident ion and can be described in 
term of the stopping power $S$, which is widely used in the treatment of ion-solid 
collisions. The stopping power represents the energy loss per unit 
length and is defined~as 
\begin{equation}\label{stopping_elec_eq}
S = \frac{{\rm d}T}{{\rm d}s} = -\,\gamma(r_{\rm s})\,v
\end{equation}
where d$T$ is the energy loss over the pathlength d$s$ \citep{sig81}. 
The total energy loss can then be obtained integrating Eq.~\ref{stopping_elec_eq}
\begin{equation}\label{Eloss_gen_eq}
T_{\rm e} = \int\,\gamma(r_{\rm s})\,v\,{\rm d}s
\end{equation}
The friction coefficient $\gamma$ \citep{puska83} depends on the density 
parameter $r_{\rm s} = \left(\frac{4}{3} \pi n_{0}\right)^{-1/3}$ which is a 
function of the valence electron density $n_{0}$. The choice of the valence 
electron density is a delicate issue for PAHs. In the case of fullerene, 
this is assumed to be the spherically symmetric jellium shell calculated by 
\citet{puska93}, which can be approximated by the following expression \citep{hadjar01}:
\begin{equation}\label{hadj_distr_eq}
n_{0} = 0.15\,\exp\,[-(r-6.6)^{2}\,/\,2.7]
\end{equation}
where $r$ (in atomic units, a.u.\footnote{Length (a.u.) = $a_0$ = 0.529 \AA, 
Velocity (a.u.) = 0.2\,$\sqrt{E({\rm keV/amu})}$, Energy (a.u.) = 27.2116 eV}) 
is the distance from the fullerene center.  The electron density
decays outside the shell and toward the center in such a way that the
major contribution comes from the region with $4 < r < 9$.

The similarity in $\pi$ electronic structure and bonding allows us to 
apply this jellium model also to PAHs. However, the spherical geometry
is clearly not appropriate for PAHs, which we model instead as a thick disk 
analogously to the distribution from Eq. \ref{hadj_distr_eq}
\begin{equation}\label{Xdistribution_eq}
n_{0} = 0.15\,\exp\,[-(x)^{2}\,/\,2.7]
\end{equation}
where $x$ is the coordinate along the thickness of the disk. This electron 
density peaks at the center of the molecule and vanishes outside, leading 
to a thickness $d\:\sim\:$4.31~$\AA$. The radius $R$ of the disk is given by 
the usual expression for the radius of a PAH: 
$a_{\rm PAH} = R = 0.9\,\sqrt(N_{\rm C})\;\AA$, where $N_{\rm C}$ is the number 
of carbon atoms in the molecule \citep{omont86}. For a 50 C-atoms PAH, 
$R = 6.36\,\AA$.


\begin{figure}
  \centering
  \includegraphics[width=.8\hsize]{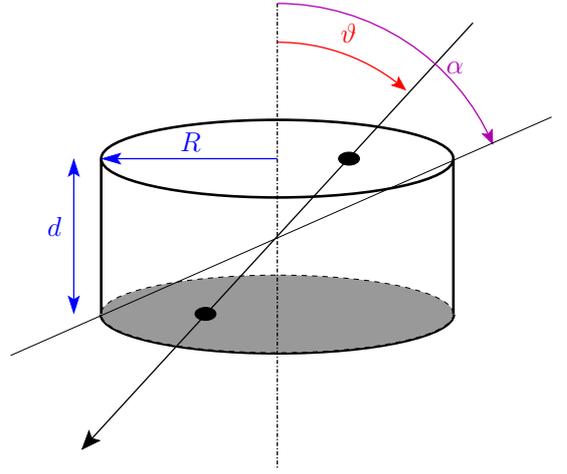}
  \caption{The coordinate system adopted to calculate the energy transferred
           to a PAH via electronic excitation by ion collisions and by impacting 
           electrons. The molecule
           is modeled as a disk with radius $R$ and thickness $d$. The trajectory
           of the incoming particle is identified by the angle $\vartheta$, while
           the angle $\alpha$ corresponds to the diagonal of the disk.
           }
  \label{pah_fig}
\end{figure}


To calculate the energy transferred from Eq. \ref{Eloss_gen_eq}, we adopt
the coordinate system shown in Fig. 1, where the pathlength through the
PAH $s$ is expressed as a function of the coordinate $x$  
and of the angle $\vartheta$ between the axes of the molecule and the 
direction of the incoming ion. In this way for each trajectory given 
by $\vartheta$ the corresponding energy transferred can be computed. 
We have d$x = $d$s\,\cos\,\vartheta$ and the electron density is then given by
\begin{equation}\label{Sdistribution_eq}
n_{0}(s,\vartheta) = 0.15\,\exp\,[-(s\,\cos\,\vartheta)^{2}\,/\,2.7\,]
\end{equation}
The density parameter can be rewritten as $r_{\rm s}(s,\vartheta) = 
\left(\frac{4}{3} \pi n_{0}(s,\vartheta)\right)^{-1/3}$
The friction coefficient $\gamma$ has been calculated by \citet{puska83} for 
various projectile ions and $r_{\rm s}$ values. It can be interpolated by the
exponential function
\begin{equation}\label{gamma_eq}
\gamma(r_{\rm s}) = \Gamma_{0}\,\exp\,\left[\frac{-(r_{\rm s}(s,\vartheta)-1.5)}{R_{2}}\right]
\end{equation}
where $\Gamma_{0}$ is the value of $\gamma$ when $r_{\rm s} = 1.5$. For hydrogen, 
helium and carbon the fit parameter $R_{2}$ equals 2.28, 0.88 and 0.90, respectively.

The energy transferred is then given by the following equation
\begin{equation}\label{Eloss_theta}
  T_{\rm e}(\vartheta) = 27.2116\,\int_{-R/\sin\,\vartheta}^{R/\sin\,\vartheta}v\,\gamma(r_{\rm s})\,{\rm d}s
\end{equation}
with $\gamma(r_{\rm s})$ from Eq. \ref{gamma_eq}. The constant 27.2116 converts the 
dimensions of $T_{\rm e}$ from atomic units to eV. For a given velocity of the incident 
particle, the value of $T_{\rm e}$ will be maximum for $\vartheta = \pi/2$ ($s = 2R$, 
the longest pathlength) and minimum for $\vartheta = 0, \pi$ ($s = d$, the shortest 
pathlength). The deposited energy also increases with the pathlength across the 
molecule, and then will be higher for larger PAHs impacted under a glancing collision. 

The energy transferred determines the PAH dissociation probability
upon electronic excitation, which is required to quantify the destruction induced
by inelastic energy loss in the hot gas (see Sect. 4).

\subsection{Nuclear interaction above threshold}

In the present study we have to consider not only the electronic interaction 
described above, but the nuclear part of the ionic collision as well.
The theory of nuclear interaction above threshold has been described in detail 
in MJT. We summarize here for clarity the essential concepts and the equations 
which will be used in the following.

For the nuclear interaction we consider only collisions above threshold, i.e. 
collisions able to transfer more than the minimum energy $T_{\rm 0}$ required 
to remove a C-atom from the PAH. To calculate the PAH destruction due to nuclear
interaction, we use the rate of collisions above threshold between PAHs and ions 
in a thermal gas, as given by Eq. 30 in MJT:
\begin{equation}\label{rate_thermalNuc_eq}
  R_{\rm n,T} = N_{\rm C}\,0.5\,\chi_{i}\,n_{\rm H}\, \int_{v_{0}}^{\infty}F_{\rm C}
               v\,\sigma(v)\,f(v,T)\,{\rm d}v
\end{equation} 
where $f(v,T)$ is the Maxwellian velocity distribution function for ion, $i$, 
in the gas, $F_{\rm C}$ is the Coulombian correction factor, $\sigma(v)$ is 
the nuclear interaction cross section per atom above threshold (Eq. 22 in MJT
where $E$ is the kinetic energy corresponding to $v$), and the factor 1/2 takes 
the averaging over the orientation of the disk into account. 

For $T_{\rm 0}$ we adopt a value of 7.5~eV (see discussion in MJT, Sect. 2.2.1). 
The kinetic energy required for the incoming ion to transfer $T_{\rm 0}$ is the 
critical energy $E_{0}$. The lower integration limit $v_{0}$ in 
Eq. \ref{rate_thermalNuc_eq} is the critical velocity corresponding to $E_{\rm 0n}$,
which is the minimum kinetic energy for the projectile to have the nuclear 
interaction cross section different from zero (cf. Sect. 2.2 in MJT).

\section{Electron collisions with PAHs}

Fast electrons are abundant in a hot gas. Because of their low mass, they 
can reach very high velocities with respect to the ions, and hence high
rates of potentially destructive collisions. We consider gas temperatures
up to 10$^{8}$ K, corresponding to a thermal electron energy of $\sim$10 keV,
well below the relativistic limit of $\sim$500 keV.
Under these low-energy conditions, with respect to the relativistic regime, 
elastic collisions between electrons and target nuclei are not effective. 
The energy transfer occurs through inelastic interactions with target electrons 
(as for electronic excitation by impacting ions), which lead to a collective 
excitation of the molecule, followed eventually by dissociation or relaxation 
through IR emission.

The calculation of the energy transferred by such `slow' electrons is in fact
a delicate matter. The theory developed under the first Born approximation
\citep{bethe30} can be applied only to the most energetic electrons (around 
few keV) but is unsuitable for the rest of our range. At low energies ($<$10 keV),
where the first Born approximation is no longer valid, the Mott elastic cross
section must be used instead of the conventional Rutherford cross section
\citep{mott49, czy90}. The Monte Carlo program CASINO \citep{hov97} computes the
Mott cross sections in the simulation of electron interactions with various
materials. Unfortunately the stopping power d$E/$d$x$ is not included in the program
output. An empirical expression for d$E/$d$x$ has been proposed by \citet{joy89}, 
which nevertheless is reliable only down to 50 eV, while we are interested in 
the region between 10 and 50 eV as well.

\begin{figure}
  \centering
  \includegraphics[width=.9\hsize]{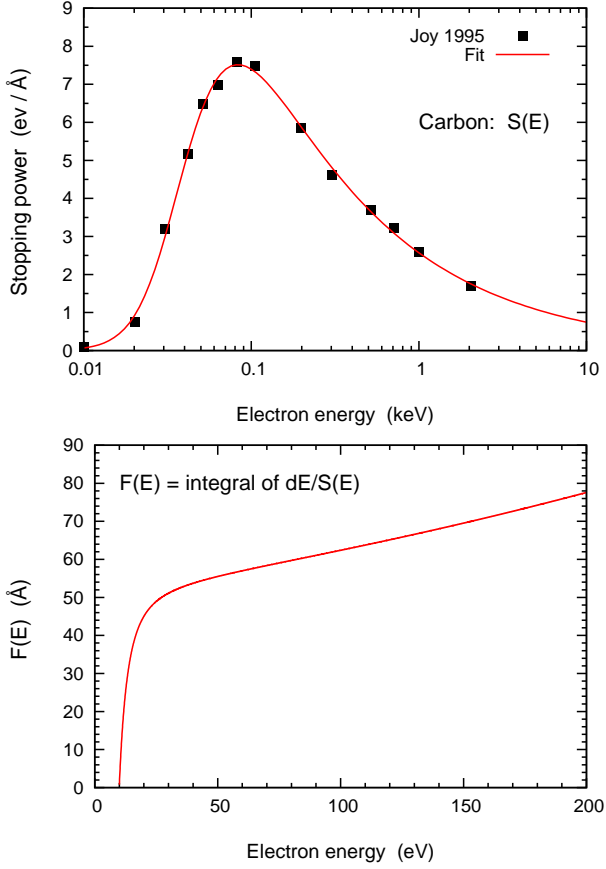}
  \caption{Top panel - Experimental measurement of d$E/$d$x$ in solid carbon 
           for electrons with energy between 10 eV and 2 keV, from \citet{joy95}, 
           overlaid with the fitting function $S(E)$ (solid line).
           Bottom panel - Integral $F$ calculated numerically as a function of 
           the energy of the incident electron, $E$.
           }
  \label{Fit_fig}
\end{figure}

We decided thus to derive the electron stopping power from experimental results.
Measurements of the electron energy loss in PAHs are not available in our energy
range of interest, so we use the measurement of d$E/$d$x$ in solid carbon for electrons
with energy between 10 eV and 2 keV \citep{joy95}. The data points are well fitted 
(to within few \%) by the following function:
\begin{equation}\label{Estopping_eq}
  S(E) = \frac{h\,\log(1+a\,E)}{f\,E^{g}\,+\,b\,E^{d}\,+\,c\,E^{e}}
\end{equation}
where $E$ is the electron energy (in keV). The values of the fitting parameters
are reported in Table~\ref{stopping_tab}.
$S(E)$ has the same functional form as the ZBL reduced stopping cross section for nuclear 
interaction (cf. Sect. 2.1 in MJT). The datapoints and the fitting function are shown in
the top panel of Fig. \ref{Fit_fig}.


\begin{table}
  \begin{minipage}[t]{\columnwidth}
    \caption
        {Analytical fit to the electron stopping power in solid carbon.
        }
        \label{stopping_tab}      
        \centering          
        \renewcommand{\footnoterule}{}      
        \begin{tabular}{c c c c}     
          \noalign{\smallskip}
          \noalign{\smallskip}
          \hline\hline       
          \noalign{\smallskip}
          \multicolumn{4}{c}{$S(E) = h\,\log(1+a\,E)\,/\,f\,E^{g}\,+\,b\,E^{d}\,+\,c\,E^{e}$}  \\ 
          \noalign{\smallskip}
          \hline
          \noalign{\smallskip}
          $a$  &  $b$  &  $c$  &  $d$   \\
          \noalign{\smallskip}
          -0.000423375  &  -3.57429$\times$10$^{-11}$  &  -3.37861$\times$10$^{-7}$  &  -3.18688  \\
          \noalign{\smallskip}
          \hline
          \noalign{\smallskip}
          $e$  &  $f$  &  $g$  &  $h$   \\
          \noalign{\smallskip}
          \noalign{\smallskip}
          -0.587928     &  -0.000232675               &   1.53851                   &  1.41476  \\
          \noalign{\smallskip}
          \hline 
        \end{tabular}        
  \end{minipage}
\end{table}      



The stopping power increases sharply at low electron energies, reaches its maximum
at $\sim$0.1 keV, and decreases smoothly afterwards. Between 0.01 and 0.1 keV a small
variation of the energy of the incident electron will translate into a large change
in the transferred energy per unit length. The shape of $S(E)$ implies that
only those electrons with energies that fall in a well defined window will efficiently
transfer energy, while electrons below $\sim$0.02 keV and above $\sim$2 keV are not
expected to contribute significantly to PAH excitation.

Once the stopping power d$E/$d$x$ is known, we can calculate the energy
transferred by an electron of given energy when travelling through the PAH. We adopt
the same configuration used for electronic interaction, shown in Fig. \ref{pah_fig}.
The trajectory of the incoming electron is defined by its impact angle $\vartheta$ and
by the geometry of the molecule. 

Because the thickness of the PAH is non-negligible with respect to its radius,
the stopping power is not constant along the electron path. To calculate the 
energy loss we thus follow the procedure described below. We have d$E/$d$x = -S(E)$,
then
\begin{equation}
  \int {\rm d}x = -\int \frac{{\rm d}E}{S(E)} = F(E)
\end{equation}
Thus, $x_{1}-x_{0} = -\left[ F(E_{1})-F(E_{0})\right]$ where $x_{1}\,-\,x_{o} = l(\vartheta)$ 
is the maximum pathlength through the PAH. From inspection of Fig. \ref{pah_fig}, 
one can see that, if $|\tan(\vartheta)| < \tan(\alpha)$, $l(\vartheta) = d/|\cos\vartheta|$, 
otherwise $l(\vartheta) = 2R/|\sin\vartheta|$. $E_{0}$ is the initial energy of the 
impacting electron, $E_{1}$ is the electron energy after having travelled the distance 
$l(\vartheta)$, which needs to be calculated so we can then determine the energy 
transferred to the PAH
\begin{equation}\label{eTransfElec_eq}
  T_{\rm elec}(\vartheta) = E_{0} - E_{1}
\end{equation}
The integral $F$, calculated numerically as a function of $E$, is shown in the bottom
panel of Fig.~\ref{Fit_fig}. We recognize that for low energies, $F$ rises sharply, 
reflecting the small energy stopping power in this energy range (cf., top panel in 
Fig.~\ref{Fit_fig}). For higher energies, $F(E)$ rises slowly (and linearly) with 
increasing energy over the relevant energy range. We note that the initial rise 
depends strongly on the (uncertain) details of the stopping power at low energies. 
However, it is of no consequence in our determination of the amount of energy deposited 
since we are only concerned with those collisions for which the energy deposition is 
in excess of the threshold energy ($\ga 10$ eV) over a pathlength of $\simeq 10-20$ \AA . 
Or to phrase it differently, at low energies a much larger pathlength has to be traversed 
(than is relevant for PAHs) in order to transfer sufficient energy to cause fragmentation.
Then, since we know $E_{0}$, we can calculate $F(E_{0}$), and $F(E_{1})$ is then given by 
$F(E_{0}) - l(\vartheta)$. We also calculated $E$ as a function of $F(E)$. We can then 
determine $E_{1}$ from $F(E_{1})$ and, finally, $T_{\rm elec}$ from Eq. \ref{eTransfElec_eq}.

\section{PAH destruction}

\subsection{Dissociation probability}

Ion or electronic collisions (or UV photon absorption) can leave the 
molecule internally (electronically) excited with an energy $T_{\rm E}$. 
Internal conversion transfers this energy to vibrational modes, and 
the molecule can then relax through dissociation or IR emission. 
These two processes are in competition with each other. To quantify 
the PAH destruction due to ion and electron collisions we need 
to determine the probability of dissociation rather than IR emission.

In the microcanonical description of a PAH, the internal energy, $T_{\rm E}$, 
is (approximately) related to the effective temperature of the system, 
$T_{\rm eff}$, by the following equation
\begin{equation}\label{teff1_eq}
  T_{\rm eff} \simeq 2000\,\left(\frac{T_{\rm e}(\rm eV)}{N_{\rm C}}\right)^{0.4}\,
  \left(1\,-\,0.2\,\frac{E_{\rm 0}(\rm eV)}{T_{\rm e}(\rm eV)}\right)
\end{equation}
where $E_{\rm 0}$ is the binding energy of the fragment \citep{tielens05}.
This effective temperature includes a correction for the finite size of 
the heat bath in the PAH. The temperature $T_{\rm eff}$ is defined through 
the unimolecular dissociation rate, $k_{\rm diss}$, written in Arrhenius form
\begin{equation}\label{unimol_eq}
  k_{\rm diss} = k_{\rm 0}\,(T_{\rm eff})\,\exp \left[-\,E_{\rm 0}/k\,T_{\rm eff}\right]
\end{equation}
where $k$ = 8.617$\times$10$^{-5}$ eV/K is the Boltzmann's constant \citep[cf.][]{tielens05}.

Consider the competition between photon emission at a rate $k_{\rm IR}$ (photons s$^{-1}$)
and dissociation at a rate of $k_{\rm diss}$ (fragments s$^{-1}$). For simplicity,
we will assume that all photons have the same energy, $\Delta \varepsilon$. The
probability that the PAH will fragment between the $n$th and ($n+1$)th photon
emission is given by
\begin{equation}\label{phi_eq}
   \varphi_{n} = p_{n+1}\,\prod_{i = 0}^{n} (1-p_{i})
\end{equation}
The (un-normalized) probability per step $p_{i}$ is given by 
$p_{i} = k_{\rm diss}(E_{i})/k_{\rm IR}(E_{i})$
and $E_{i} = (T_{\rm E}-i\times \Delta \varepsilon$), with $T_{\rm E}$ the 
initial internal energy, coincident with the energy transferred into the PAH. 
The total dissociation probability is then given by
\begin{equation}\label{totalProb_eq}
  P(n_{\rm max}) = \sum_{n = 0}^{n_{\rm max}} \varphi_{n}
\end{equation}
If we ignore the dependence of $k_{\rm IR}$ on $E_{i}$, the temperatures
(Eq.~\ref{teff1_eq}) drop approximately by 
$T_{i}/T_{i-1} = (1-0.4 \Delta \varepsilon/T_{\rm E})$.
The probability ratio is approximately:
\begin{equation}\label{pRatio_eq}
  \frac{p_{i}}{p_{i-1}} = \exp \left[ \left(\frac{E_{\rm 0}}{kT_{i-1}}\right)\,
    0.4\,\frac{\Delta \varepsilon}{\varepsilon} \right]
\end{equation}
These equations become very difficult to solve in closed form. However, let us just
assume that $p_{i}$ does not vary and is given by $p_{\rm av}$. Then we have the
total un-normalized dissociation probability
\begin{equation}\label{totalProb2_eq}
  P(n_{\rm max}) = (n_{\rm max}+1)\,p_{\rm av} = \frac{k_{\rm 0}\,
    \exp \left[-E_{\rm 0}/k\,T_{\rm av} \right]}{k_{\rm IR}/(n_{\rm max}+1)}
\end{equation}
where we adopt $k_{\rm IR}$ = 100 photons s$^{-1}$ \citep{joc94}, and the average
temperature is chosen as the geometric mean 
\begin{equation}\label{Tav_eq}
  T_{\rm av} = \sqrt{T_{\rm 0}\times T_{n_{\rm max}}}
\end{equation}
where $T_{\rm 0}$ and $T_{n_{\rm max}}$ are the effective temperatures corresponding to $T_{\rm E}$
(initial internal energy equal to the energy transferred) and 
$(T_{\rm E} - n_{\rm max} \times \Delta \varepsilon)$
(internal energy after the emission of $n_{\rm max}$ photons).
For the energy of the emitted IR photon we adopt the value $\Delta \varepsilon$ = 0.16 eV,
corresponding to a typical CC mode.

If the $p_{i}$'s were truly constant, then $n_{\rm max}$ would be 
$n_{\rm max}$ = $(T_{\rm E}-E_{\rm 0})/\Delta \varepsilon$.
However, they do decrease. So, rather we take it to be when the probability per step has
dropped by a factor 10. A direct comparison between the full evaluation and this simple
approximation yields $n_{\rm max}$ = 10, 20 and 40 for $N_{\rm C}$ = 50, 100 and 200
respectively. The quantity $n_{\rm max}$ scales with $N_{\rm C}$ because for a constant
temperature (eg., required to get the dissociation to occur), the internal energy has to
scale with $N_{\rm C}$. As a result, the number of photons to be emitted also has to scale
with $N_{\rm C}$.

\begin{figure}
  \centering
  \includegraphics[width=.8\hsize]{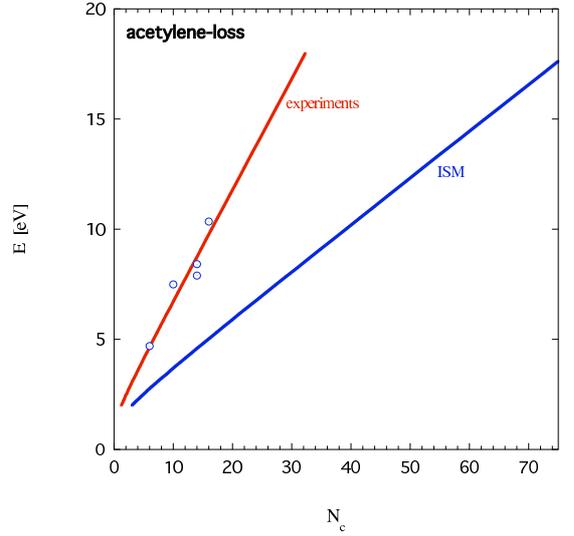}
  \caption{The appearance energy as a function of the number of C-atoms 
           in the PAH. The red line provides a fit to the experimental 
           data using Eq.~\ref{unimol_eq} for an assumed pre-exponential 
           factor $k_0= 1.4\times 10^{16}$ and a (fitted) Arrhenius energy, 
           $E_0=3.65$ eV. The data points are the experimental results of 
           \citet{joc94}. The blue line is the appearance energy for ISM
           conditions (eg., at a rate of 1 s$^{-1}$).
           }
  \label{Eapp_fig}
\end{figure}

The choice of the values to adopt for $k_{\rm 0}$ and $E_{\rm 0}$ is a delicate matter. In the 
laboratory the dissociation of highly vibrationally excited PAHs is typically measured on 
timescales of 1-100 $\mu$s because either the molecules are collisionally de-excited 
by ambient gas or the molecules have left the measurement zone of the apparatus. In contrast, 
in the ISM, the competing relaxation channel is through IR emission and occurs typically on a 
timescale of 1 s. As is always the case for reactions characterized by an Arrhenius law, a 
longer timescale implies that the internal excitation energy can be lower. This kinetic shift 
is well established experimentally and can amount to many eV. Moreover, only small PAHs (up to 
24 C-atoms) have been measured in the laboratory and the derived rates have to be extrapolated 
to much larger ($\sim$ 50 C-atoms) PAHs that are astrophysically relevant. In an astrophysical 
context, the unimolecular dissociation of highly vibrationally excited PAHs -- pumped by FUV 
photons -- has been studied experimentally by \citet{joc94} and further analyzed by \citet{lePage01}. 
Here, we will modify the analysis of \citet{tielens05} for H-loss by UV pumped PAHs to determine 
the parameters for carbon loss. The dissociation rate -- given by Eq.~\ref{unimol_eq} -- 
is governed by two factors, the pre-exponential factor $k_{\rm 0}$ and the energy $E_{\rm 0}$. 
The pre-exponential factor is given by 
\begin{equation} \label{pre-exp}
  k_{\rm 0}\, =\, \frac{kT_{\rm eff}}{h}\, \exp\left[ 1\, +\, \frac{\Delta S}{R}\right]
\end{equation}
where $\Delta S$ is the entropy change which we set equal to 10 cal mole$^{-1}$ 
\citep{ling98}. We will ignore the weak temperature dependence 
of $k_{\rm 0}$ in the following and adopt $1.4\times 10^{16}$ s$^{-1}$. The parameter $E_{\rm 0}$ 
can now be determined from a fit to the experiments by \citet{joc94} on small PAHs. These 
experiments measured the appearance energy $E_{\rm ap}$ -- the internal energy at which noticable 
dissociation of the PAH first occurred. The rate at which this happened was assumed to be 
$10^{4}$ s$^{-1}$. Adopting this value, the appearance energy can be estimated from Eq.~\ref{unimol_eq}. 
The results of our analysis are shown in Fig. \ref{Eapp_fig}. The internal energy required 
to dissociate a PAH depends strongly on the PAH size. Likewise the kinetic shift associated 
with the relevant timescale at which the experiment was performed is quite apparent. Indeed, 
this kinetic shift can amount to tens of eV for relevant PAHs.


The derived Arrhenius energy of 3.65 eV is small compared to the binding energy 
of a C$_{\rm 2}$H$_{\rm 2}$ group in a PAH \citep[4.2 eV,][]{ling98}. This is a well known 
problem in statistical unimolecular dissociation theories \citep[cf.][]{tielens08}. 
We emphasize that these results show that a typical interstellar PAH with a size 
of 50 C-atoms would have a dissociation probability of $\sim 1/2$ after absorption 
of an FUV photon of $\sim$12 eV (cf. Fig. \ref{Eapp_fig}). Hence, PAHs would be 
rapidly lost in the ISM through photolysis. It seems that the experiments on small 
PAHs cannot be readily extrapolated to larger, astrophysically relevant PAHs. 
Possibly, this is because experimentally C$_2$H$_2$ loss has only been observed 
for very small catacondensed PAHs with a very open carbon skeleton (e.g., 
naphthalene, anthracene, and phenanthrene) which are likely much more prone 
to dissociation than the astrophysically more relevant pericondensed PAHs. 
Indeed, the small pericondensed PAHs, pyrene and coronene did not show any 
dissociation on the experimental timescales \citep{joc94, ling98}.


\begin{figure}
  \centering
  \includegraphics[width=.8\hsize]{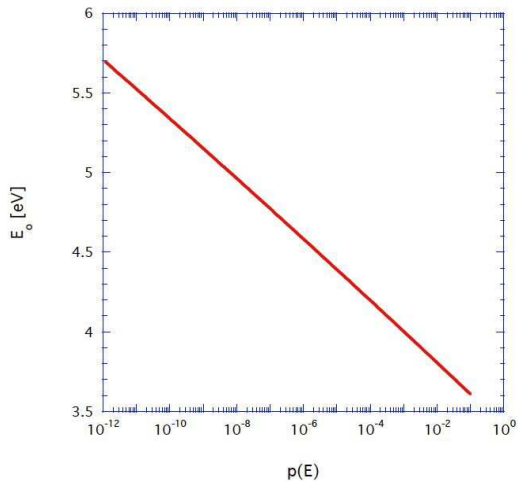}
  \caption{The probability for dissociation of a 50 C-atom PAH excited by 
           10 eV as a function of the Arrhenius energy, $E_{\rm 0}$.
           }
  \label{Eo_fig}
\end{figure}


Turning the problem around, we can determine the Arrhenius energy, $E_{\rm 0}$, 
as a function of the dissociation probability by adopting an IR relaxation rate 
of 1 s$^{-1}$ and an internal excitation energy equal to a typical FUV photon 
energy (12 eV). The results for a 50 C-atom PAH are shown in Fig \ref{Eo_fig}. 
If we adopt a lifetime, $\tau_{\rm PAH}$, of 100 million years, a PAH in the 
diffuse ISM will have typically survived some 
$\sigma_{\rm uv} \, N_{\rm uv}\, \tau_{\rm PAH} =\, 2\times 10^{6}\, N_{\rm C}$ UV 
photon absorptions (with $\sigma_{\rm uv}=7\times 10^{-18}$ cm$^2$, 
$N_{\rm uv}=10^8$ photons cm$^{-2}$ s$^{-1}$ in a Habing field). Hence, if the 
lifetime of the smallest PAHs in the ISM (eg., with $N_{\rm C}\simeq 50$ C-atom) 
is set by photodissociation of the C-skeleton, the probability for dissociation 
has to be $5\times 10^{-7}$ corresponding to an Arrhenius energy of 4.6 eV 
(Fig. \ref{Eo_fig}). We note that in a PDR the photon flux is higher 
($G_{\rm 0}\sim 10^{5}$) while the lifetime (of the PDR) is smaller 
($\tau_{\rm PDR}\simeq 3\times 10^{4}$ yr), resulting in $6\times 10^{7}$ UV 
photons absorbed over the PDR lifetime. Survival of PAHs in a PDR environment 
would therefore require a somewhat larger $E_0$ (or alternatively, only slightly 
larger PAHs could survive in such an environment).

We note that the binding energy of a C$_{\rm 2}$H$_{\rm 2}$ group to small PAHs is 
estimated to be 4.2 eV and is probably somewhat larger for a 50 C-atom condensed 
PAH. Loss of pure carbon, on the other hand, requires an energy of 7.5 eV, close 
to the binding energy of C to graphite. Loss of C$_{\rm 2}$ from fullerenes has a 
measured $E_{\rm 0}$ of $9.5\, \pm\, 0.1$ eV \citep{tomita01}. These latter two 
unimolecular dissociation channels are for all practical purposes closed under 
interstellar conditions. Finally, likely, H-loss will be the dominant destruction 
loss channel for large PAHs ($E_0=3.3$ eV), leading to rapid loss of all H's \citep{lePage01, 
tielens05}. The resulting pure C-skeleton may then isomerize to much more stable 
carbon clusters, in particular fullerenes, and this may be the dominant `loss' 
channel for interstellar PAHs \citep[cf.][ and references therein]{tielens08}. 
It is clear that there are many uncertainties in the chemical destruction routes 
of interstellar PAHs and that these can only be addressed by dedicated experimental 
studies. For now, in our analysis of the unimolecular dissociation of PAHs -- excited 
by electron or ion collisions -- we will adopt $E_{\rm 0}=4.6$ eV as a standard value. 
We will however also examine the effects of adopting $E_{\rm 0}=3.65$ eV (eg., $p=1/2$) 
and $E_{\rm 0}=5.6$ eV (eg., $p=3\times 10^{-12}$).

\subsection{Collision rate}

Once the dissociation probability is determined, we can calculate the destruction 
rate through electronic excitation following electron or ion collision. Adopting the 
configuration shown in Fig. \ref{pah_fig}, the destruction rate is given by
\begin{equation}\label{collRateElec0_eq}
  R_{\rm e} = v\,\chi\,n_{\rm H}\,F_{\rm C}\,\frac{1}{2\pi}
                         \int {\rm d}\Omega\,\sigma_{\rm g}(\vartheta)\,
                         P(v,\vartheta)
\end{equation}
with $\Omega = \sin \vartheta {\rm d}\vartheta {\rm d}\varphi$, $\varphi$ running from 0 to 2$\pi$ 
and $\vartheta$ from 0 to $\pi/2$. Eq. \ref{collRateElec0_eq} can be re-written as
\begin{equation}\label{collRateElec_eq}
  R_{\rm e} = v\,\chi\,n_{\rm H}\,F_{\rm C}
                         \int_{\vartheta=0}^{\pi/2}\sigma_{\rm g}(\vartheta)\,
                         P(v,\vartheta)\,\sin\vartheta\,{\rm d}\vartheta
\end{equation}
The term $v$ is the velocity of the incident particle, $\chi$ is the ion/electron 
relative abundance in the gas, and $F_{\rm C}$ is the Coulombian correction factor,
taking into account the fact that both target and projectiles are charged, and thus the
collision cross section could be enhanced or diminished depending on the charge sign.
The electron coulombian factor is always equal to 1 (to within 1\%) because 
electrons have low mass. For a fixed temperature they reach higher velocities, 
with respect to the ions, and are thus less sensitive to the effects of the 
Coulombian field. The quantity $\vartheta$ is the angle between the vertical axes 
of the PAH and the direction of the projectile. The term $\sigma_{\rm g}$ 
is the geometrical cross section seen by an incident particle with direction defined by 
$\vartheta$. The PAH is modelled as a thick disk with radius $R$ and thickness $d$. 
The cross section is then given by
\begin{equation}\label{disk_cs_eq}
  \sigma_{g} = \pi\,R^{2}\,\cos\vartheta\,+\,2\,R\,d\,\sin\vartheta
\end{equation}
which reduces to $\sigma_{g} = \pi R^{2}$ for $\vartheta$ = 0 (face-on impact) and to
$\sigma_{g} = 2Rd$ for $\vartheta = \pi/2$ (edge-on impact).
The term $P(v,\vartheta)$ represents the total probability for dissociation
upon electron collisions and electronic excitation, for a particle with velocity
$v_{\rm PAH}$ and incoming direction $\vartheta$. It was calculated from
Eq.~\ref{totalProb2_eq} using the appropriated value of the energy transferred:
$T_{\rm elec}$ for electrons (Eq. \ref{eTransfElec_eq}) and $T_{\rm e}$ for electronic 
excitation (Eq.~\ref{Eloss_theta})

We are considering a hot gas and therefore we are interested in the thermal
collision rate, given by
\begin{equation}\label{collRateElecThermal_eq}
 R_{\rm e,T} = \int_{v_{\rm 0,e}}^{\infty}R_{\rm e}(v)\,f(v,T)\,{\rm d}v
\end{equation}
where $v_{\rm 0,e}$ is the electron/ion velocity corresponding to the electronic
dissociation energy $E_{0}$. 
The temperature $T$ is the same for both ions and electrons, but these latter will 
reach much larger velocities. From Eq. \ref{collRateElec_eq} we then expect to find 
a significantly higher collision rate with respect to the ion case.

\section{Results}

\subsection{PAH destruction in a hot gas}

To describe the destructive effects on PAHs of collisions with ions $i$ 
($i$ = H, He, C) and electrons in a thermal gas we have to evaluate the rate 
constant for carbon atom loss.

For all processes the rate constant $J$ is defined by
the ratio $J = R_{\rm T}/n_{\rm H}$, where $R_{\rm T}$ is the thermal 
collision rate appropriate for nuclear and electronic excitation and electron 
collisions (cf. Eq. \ref{rate_thermalNuc_eq} and \ref{collRateElecThermal_eq}),  
and $n_{\rm H}$ is the density of hydrogen nuclei. For electronic and electron
interactions the rate must be multiplyied by a factor of 2, to take into account
the fact that each collision leads to the loss of two carbon atoms.

\begin{figure*}
  \centering
  \includegraphics[width=.8\hsize]{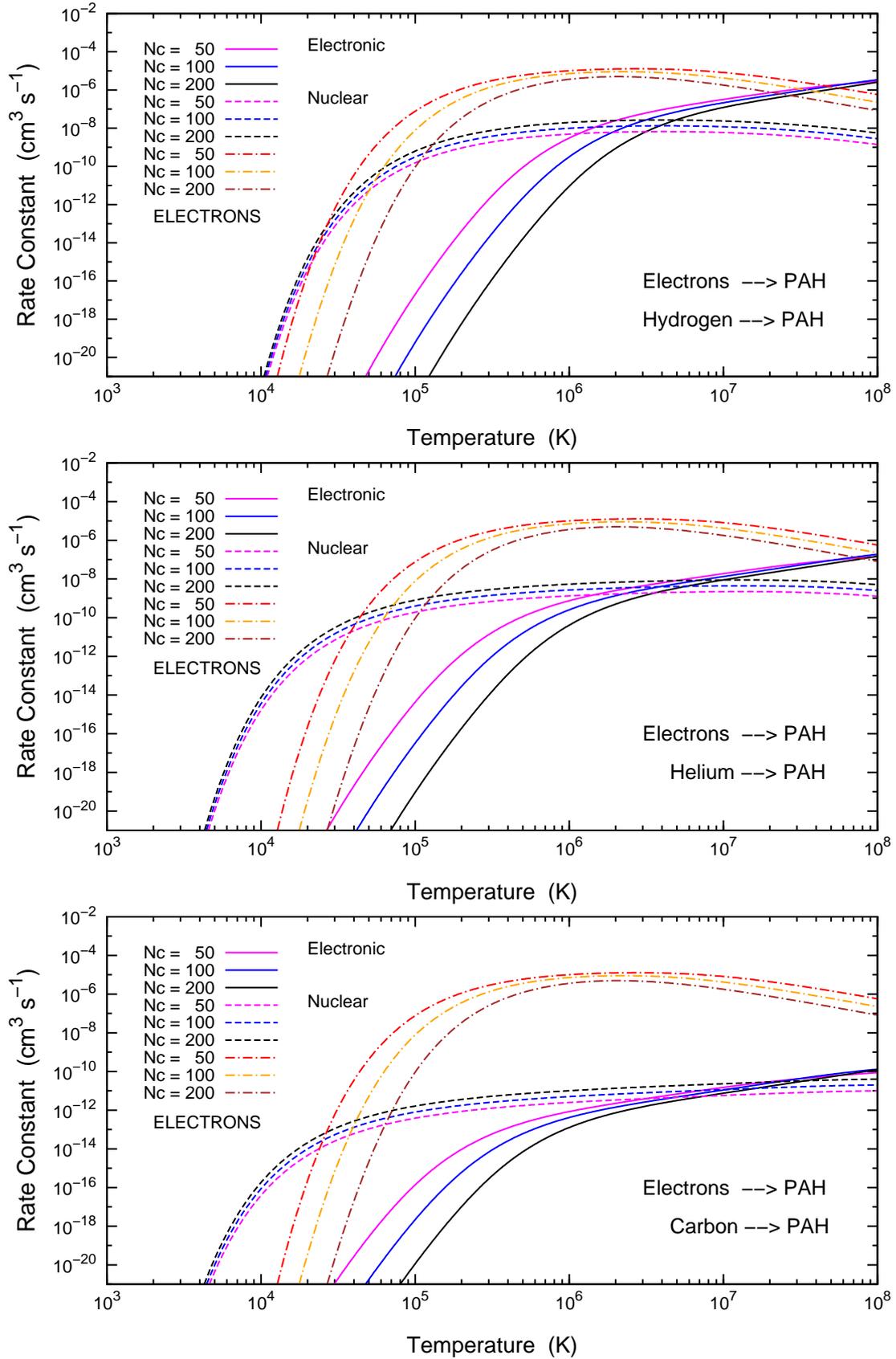}
  \caption{Nuclear (dashed lines), electronic (solid lines) and 
           electron (dashed-dotted lines) rate constant for PAH 
           carbon atom loss due to collisions with
           H, He, C and electrons in a thermal gas. The rate constants are 
           calculated as a function of the gas temperature for three
           PAH sizes $N_{\rm C}$ = 50, 100, 200 C-atoms, assuming the
           nuclear threshold energy $T_{0}$ = 7.5 eV and the electronic
           dissociation energy $E_{0}$ = 4.58 eV.
           }
  \label{rateConstant_fig}
\end{figure*}

The electron, nuclear and electronic rate constants for three PAH sizes 
(Nc = 50, 100 and 200 C-atom) are shown in Fig. \ref{rateConstant_fig} 
as a function of the gas temperature. We assume for the nuclear threshold
energy $T_{0}$ the value of 7.5 eV (MJT), and for the electronic dissociation
energy the value 4.58 eV (Sect. 4.1).

For the nuclear interaction, the threshold energy $T_{\rm 0}$ is independent 
from the PAH size, so the three curves start at the same temperature (not 
shown in the plot). The small separation between the curves is due to the 
fact that the PAH `surface area' -- and hence the rate constant -- scales
linearly with $N_{\rm C}$, therefore $J$ is higher for bigger PAHs.

For nuclear (and electronic) interactions, the rate constants decrease
from hydrogen to carbon because of the lower abundance of the heavier 
projectiles with respect to hydrogen (H : He : C = 1 : 0.1 : 10$^{-4}$).
On the other hand, the nuclear curves shift toward lower temperatures from 
lighter to heavier projectiles. This is the reflection of the fact that the 
critical energy of the particle, required to transfer the threshold energy 
$T_{0}$, decreases
\clearpage
%
%
\clearpage
%
%
\begin{figure*}
  \centering
  \includegraphics[width=.8\hsize]{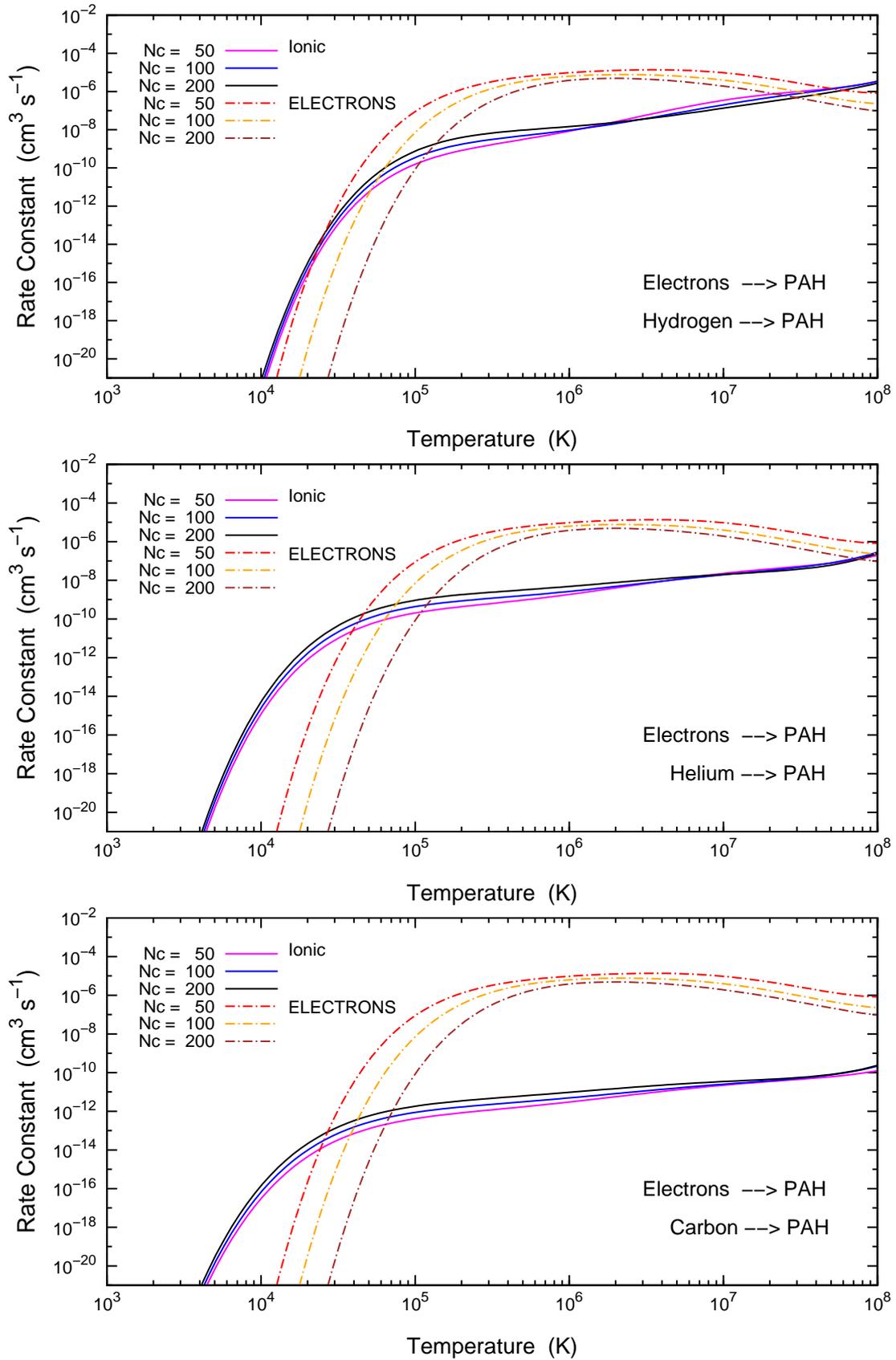}
  \caption{Analytical fits to the calculated rate constants shown in 
           Fig. \ref{rateConstant_fig}. The `Ionic' curve is the fit 
           to the sum of the nuclear and electronic rate constants,
           thus represents the total contribution from ion collisions
           to PAH destruction. 
           }
  \label{rateConstantFit_fig}
\end{figure*}
%
\clearpage
\noindent
with increasing mass of the projectile itself. Then, a 
carbon atom with a temperature of $\sim$ 4$\times$10$^{3}$ K is hot enough 
to transfer the energy required for atom removal via nuclear interaction, 
whereas for hydrogen a temperature of at least 10$^{4}$ K is necessary. 
The almost-constant behaviour after the initial rise reflects the large 
maximum observed in the nuclear cross section (cf. Fig. 2 in MJT).

The dissociation probability $P(n_{\rm max})$ (Eq. \ref{totalProb2_eq}) depends on 
the binding energy of the fragment $E_{0}$, on the PAH size $N_{\rm C}$ and on the 
energy transferred (through $T_{\rm av}$), which in turns depends on the initial energy 
(velocity) of the projectile. For a thermal distribution, this latter will be 
determined by the gas temperature~$T$. 

The electronic rate constant curves for the different PAH sizes are well separated 
at the lowest temperatures. This reflects the fact that, for a fixed value of the 
transferred energy and of the electronic dissociation energy $E_{0}$, the dissociation 
probability decreases for increasing $N_{\rm C}$ because either more energy is required 
in the bond that has to be broken or because the energy is spread over more vibrational 
modes and hence the internal excitation temperature is lower. On the other hand, the more 
energy that is deposited into the PAH, the higher is the dissociation probability. The energy 
transferred via electronic excitation (and then $T_{\rm av}$) increases with the energy of 
the projectile up to a maximum value, corresponding to an incident energy of $~$100 keV 
for H (and higher for more massive particles), and decreases beyond that for higher energies. 
The energy content of a thermal gas at $T$ = 10$^{8}$ K is $\sim$9 keV, thus, in the 
temperature range considered in this study, the energy transferred increases with 
temperature (energy) and hence the dissociation probability increases as well. This 
is the basis for the monotonic rise of the electronic rate constant. After the initial 
separation, the three curves converge, because the rise in the transferred energy 
compensates the effect of increasing $N_{\rm C}$. 

As discussed in Sect. 3, the energy transferred by impacting electrons rises sharply for 
energies in excess of 10 eV, peaks at $\sim$ 100 eV and decreases more slowly down to 
10 keV. This results in a dissociation probability shaped as a step function: for 
10 eV $\la v \la$ 4 keV, $P(n_{\rm max})$ jumps from values close to zero up to 1. 
These limiting energies apply to a 50 C-atom PAH; for $N_{\rm C}$ = 200 the width of 
the step is smaller (100 eV $\la v \la$ 1 keV), due to the fact that for a bigger PAH  
more energy has to be transferred for dissociation. This behaviour is reflected in the 
shape of the electron rate constant, where a steep rise is followed by a maximum, which
is emphasised by the logarithmic scale used for the plot.
As expected, the electron rate constant overcomes the electronic one, except for the
highest gas temperatures. This results from the fact that, for a given temperature,
electrons can reach higher velocities with respect to the ions (cf. Eq. 
\ref{collRateElec_eq} and \ref{collRateElecThermal_eq}). 

To summarize, from Fig. \ref{rateConstant_fig} we can infer that, according 
to our model, the destruction process is dominated by nuclear interaction with 
helium at low temperatures (below $\sim$3$\times$10$^{4}$ K), and by electron
collisions above this value. Small PAHs are easier destroyed than big ones for 
temperatures below $\sim$10$^{6}$~K, while the difference in the destruction
level reduces significantly for hotter gas.

\begin{figure*}
  \centering
  \includegraphics[width=.8\hsize]{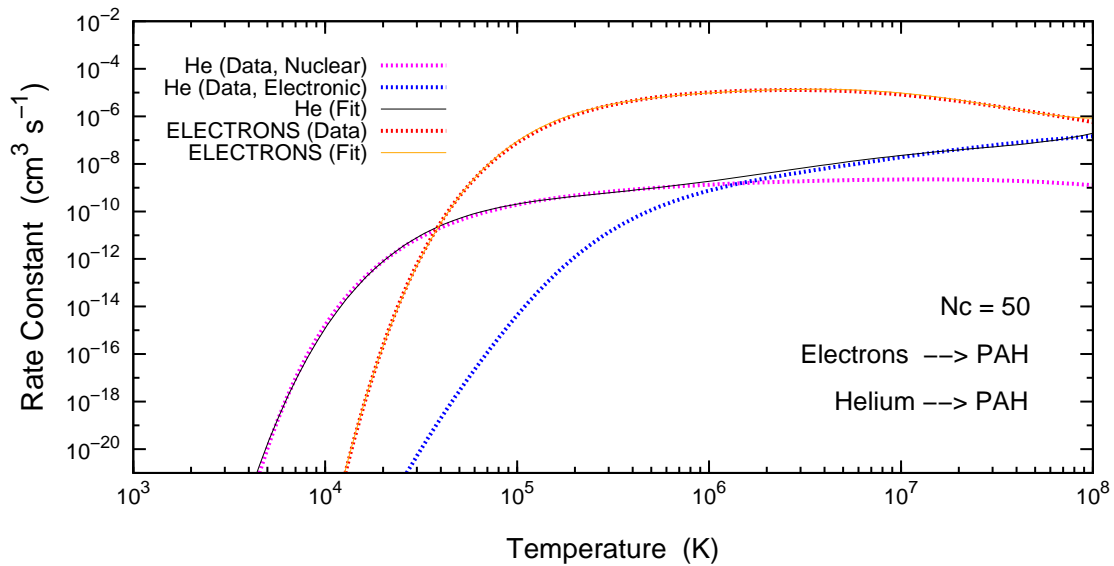}
  \caption{Calculated rate constants for electrons and helium impacting
           on a 50 C-atom PAH, overlaid are the corresponding
           analytical fits. The He fit is for the sum of the nuclear
           and electronic rate constants. The average discrepancy
           is $\sim$15 \%. 
           }
  \label{rateFit_VS_data_fig}
\end{figure*}

The calculated rate constants, shown in Fig. \ref{rateConstant_fig}, are well 
fitted by the function $g(T) = 10^{f(\log(T))}$, where $f(T)$ is a polynomial of order 5
\begin{equation}\label{fit_f_eq}
  f(T) = a+b\,T+c\,T^{2}+d\,T^{3}+e\,T^{4}+f\,T^{5} 
\end{equation}
For each PAH size we provide the fit to the electron rate constant; for the ions,
the fit is over the sum of the nuclear and electronic rate constants, in order
to provide an estimate of the global contribution from ionic collisions. The fits are
shown in Fig. \ref{rateConstantFit_fig}, and the fitting parameters are reported in 
Table \ref{fitRate_tab}. To provide an example of the accuracy of our fitting procedure, 
Fig.~\ref{rateFit_VS_data_fig} shows the comparison between the calculated rate constant 
and the corresponding analytical fit, for electrons and helium impacting on a 50 C-atom
PAH. The He fit is over the sum of the nuclear and electronic rate constants. The
average fitting discrepancy is $\sim$15 \% and the fits are therefore well within
any uncertainties in the method.

\begin{figure*}
  \centering
  \includegraphics[width=.8\hsize]{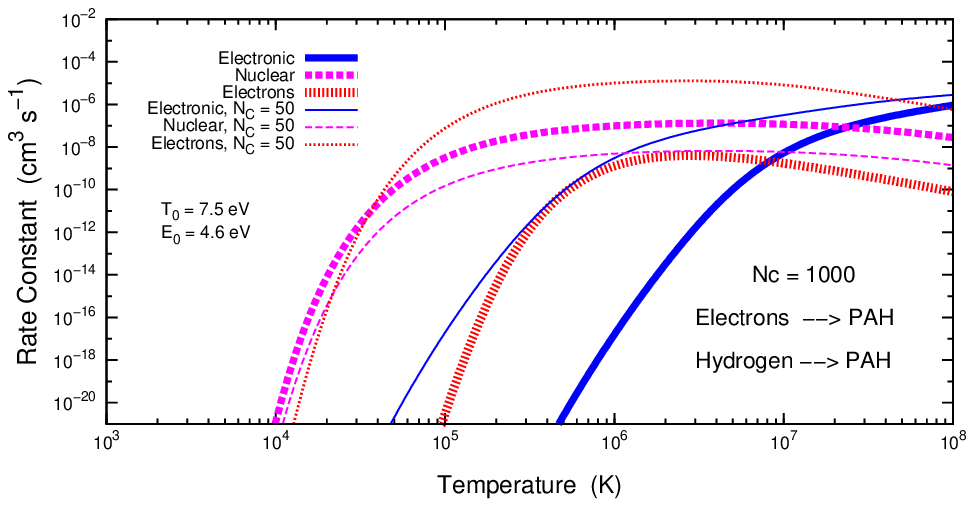}
  \caption{Carbon atom loss rate constant for electrons and hydrogen impacting
           against a 1000 C-atom PAH. The rate constants for $N_{\rm C}$ = 50
           are shown for comparison.
           }
  \label{rateConstant1000_fig}
\end{figure*}

Fig. \ref{rateConstant1000_fig} shows the comparison between the carbon loss rate 
constants for a very big PAH, $N_{\rm C}$ = 1000, and a 50 C-atom molecule. We
assume $T_{\rm 0}$ = 7.5 eV and $E_{\rm 0}$ = 4.6 eV. Because of the decrease of the
dissociation probability when the PAH size increases, as expected, the electron and 
electronic rate constants are strongly suppressed, and both curves shift toward higher 
temperatures. Indeed, for such a big PAH, a much higher internal energy is required 
to reach the internal temperature where dissociation sets in.
On the other hand, the nuclear rate
constant increases linearly with the PAH size. As a result, for a 1000 carbon atoms 
PAH, the nuclear interaction is the dominant (and efficient) destruction mechanism up to
$T \sim$ 2$\times$10$^{7}$~K. In conclusion, electrons are responsible for the
destruction of small/medium size PAH, while for big molecules this role is taken 
by ions.


%
\begin{table*}
\caption{Analytical fit parameters for the PAH carbon atom loss rate constant, 
         calculated for electron and ion collisions.}
\label{fitRate_tab}
\centering
\begin{tabular}{l c c c c c c c}     
\hline\hline
\noalign{\smallskip}
\multicolumn{2}{c}{Fitting function}  &\multicolumn{6}{c}{$f(T) = a+b\,T+c\,T^{2}+d\,T^{3}+e\,T^{4}+f\,T^{5}$}  \\ 
\noalign{\smallskip}
\hline
\noalign{\smallskip}
              &    &  $a$  &  $b$  &  $c$  &  $d$  &  $e$  &  $f$  \\
\noalign{\smallskip}
\hline
\noalign{\smallskip}
                 &  Electrons   &  2136.83  & 1632.17   & -499.822  & 76.4347 & -5.82964 & 0.177174  \\
                 &     H        &  -1896.69 & 1480.8    & -462.733  & 71.8957 & -5.54719 & 0.169996  \\
$N_{\rm C}$ = 50  &     He       &  -971.448 & 770.259   & -245.561  & 38.8995 & -3.05787 & 0.0954303 \\
                 &     C        &  -704.392 & 551.313   & -175.063  & 27.6506 & -2.16871 & 0.0675643 \\
\noalign{\smallskip}  
\hline
\noalign{\smallskip}
                 &  Electrons   &  -2255.38 & 1681.45   & -503.451  & 75.4339 & -5.64956 & 0.168959  \\
                 &     H        &  -1645.64 & 1257.06   & -384.309  & 58.4103 & -4.41007 & 0.132356  \\
$N_{\rm C}$ = 100 &     He       &  -945.901 & 747.921   & -237.984  & 37.6808 & -2.96613 & 0.0928852 \\ 
                 &     C        &  -711.244 & 558.48    & -177.983  & 28.2547 & -2.23145 & 0.0701374 \\      
\noalign{\smallskip}
\hline
\noalign{\smallskip}
                 &  Electrons   & -2234.37  & 1597.71   & -459.647  & 66.332  & -4.79841 & 0.139019  \\
                 &     H        &  -1473.64 & 1109.01   & -334.292  & 50.1417 & -3.74133 & 0.111164  \\
$N_{\rm C}$ = 200 &     He       &  -963.188 & 765.639   & -245.054  & 39.0745 & -3.10123 & 0.0980047 \\
                 &     C        & -738.791  & 584.928   & -187.884  & 30.0819 & -2.39748 & 0.0760639 \\ 
\noalign{\smallskip}
\hline
\end{tabular}
\end{table*}

\begin{figure*}
  \centering
  \includegraphics[width=.8\hsize]{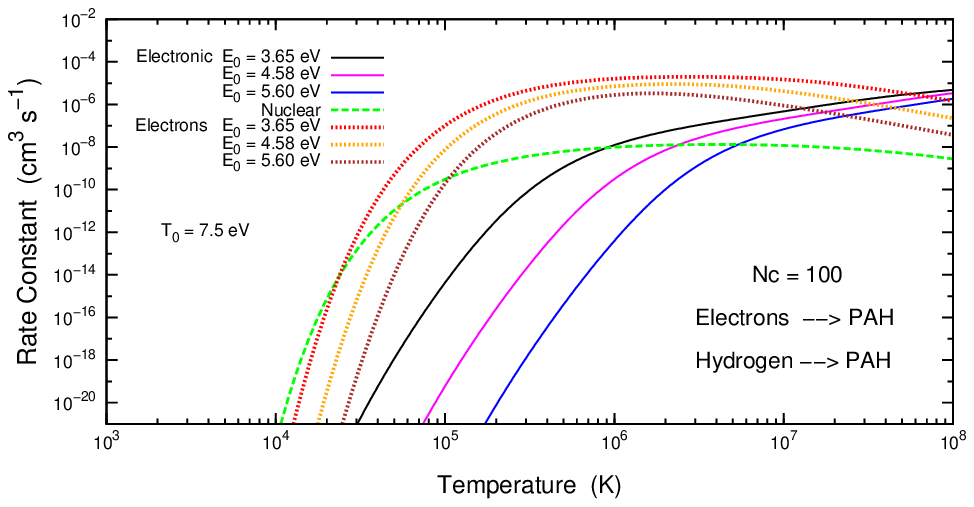}
  \caption{Comparison between carbon atom loss rate constants calculated
           assuming three values for the electronic dissociation energy, 
           $E_{\rm 0}$: 3.65, 4.58 and 5.6 eV. The curves refer to electrons 
           and hydrogen impacting against a 100 C-atom PAH. The nuclear 
           rate constant is calculated assuming $T_{\rm 0}$ = 7.5 eV.
           }
  \label{rateConstantEo_fig}
\end{figure*}

As mentioned at the end of Sect. 4.1, we examined the effects 
of adopting different
values for the Arrhenius energy, $E_0$ = 3.65 and 5.6 eV, lower and higher, 
respectively, than our canonical value 4.6 eV. The results are shown in 
Fig.~\ref{rateConstantEo_fig}. The dissociation probability decreases for
increasing $E_0$, because more energy is required in the bond that has to be
broken. Hence, as expected, both the electron and electronic rate constants
decrease in absolute value and shift toward highest temperatures. In particular
the electronic thermal shift is very pronounced, indicating how sensitive this 
process is with respect to the assumed $E_0$. A variation in the adopted electronic
excitation energy translates into a significant variation of the rate constant,
reemphasizing the importance of experimental studies of this critical energy.


%

\subsection{PAH lifetime}

Under the effect of electron and ion collisions in a hot gas, the number of 
carbon atoms in a PAH molecule varies with time. After a time $t$, this number is
\begin{equation}\label{Nt_eq}
  N_{\rm C}(t)  = N_{\rm C}(0) \exp \left[-t\,/\,\tau_{\rm 0}\right]
\end{equation}
and the number of carbon atoms ejected from this PAH is
\begin{equation}\label{Fl_eq}
  F_{\rm L}(t) = \left(1 - \exp \left[ -t\,/\,\tau_{\rm 0} \right] \right)
\end{equation}
The quantity $\tau_{\rm 0}$ is the time constant appropriate for
electron, nuclear and electronic interaction, given by
\begin{equation}\label{timeConstant_eq}
  \tau_{\rm 0}  = \frac{N_{\rm C}}{(2)\,R_{\rm T}}
               = \frac{N_{\rm C}}{J\,n_{\rm H/e}}
\end{equation}
where $R_{\rm T}$ and $J$ are the thermal collision rate and the rate constant 
for electrons, nuclear and electronic interactions respectively, and $n_{\rm H/e}$
is the hydrogen/electron density. For electrons and electronic excitation, 
the rate must be multipliyed by a factor of 2, because each interaction 
leads to the removal of two carbon atoms from the PAH. The nuclear rate 
scales linearly with $N_{\rm C}$, hence the corresponding fractional carbon 
atom loss $F_{\rm L}$ is independent of PAH size.

For any given incoming ion and fixed PAH size $N_{\rm C}$, $F_{\rm L}$
is univocally determined by the hydrogen/electron density $n_{\rm H/e}$ 
and the gas temperature $T$.  We assume that a PAH is destroyed after
the ejection of 2/3 of the carbon atoms initially present in the molecule. 
This occurs after the time $\tau_{\rm 0}$ (Eq. \ref{timeConstant_eq}) 
which we adopt as the PAH lifetime against electron and ion bombardment 
in a gas with given density and temperature.

Table \ref{object_tab} summarizes relevant data and PAH lifetime,
$\tau_{\rm 0}$, for four objects characterized by warm-to-hot gas,
X-ray emission, and (bright) IR emission features.  The lifetime has
been calculated for PAHs with $N_{\rm C}$ = 200. $\tau_{0}$(ref)
has been evaluated adopting our reference values for the interaction
parameters, $E_0$ = 4.6 eV and $T_0$ = 7.5 eV, $\tau_{0}$(min)
corresponds to the minimum values $E_0$ = 3.65 eV and $T_0$ = 4.6 eV,
and $\tau_{0}$(max) to the maximum values $E_0$ = 5.6 eV and $T_0$
= 15 eV. $\tau_{\rm object}$ is the lifetime of the object.

\begin{figure}
  \centering
  \includegraphics[width=\hsize]{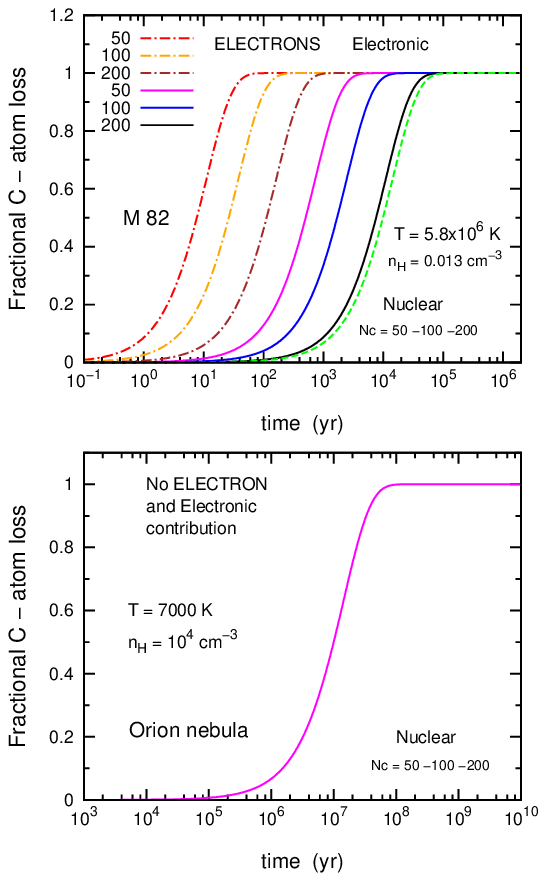}
  \caption{
           Nuclear and electronic fractional C-atom loss as a function
           of time in the M 82 galaxy (top) and in the Orion Nebula 
           (bottom), calculated for three  PAH sizes $N_{\rm C}$ = 
           50, 100, 200 C-atoms. The fractional carbon loss is defined
           as the number of carbon atoms ejected from the PAH after a 
           given time divided by $N_{\rm C}$, and represents a direct
           indicator of the level of destruction of the PAH. Each curve 
           shows the total contribution from collisions with H, He and C.
           In the nuclear case the fractional loss is independent
           from $N_{\rm C}$, so the three curves coincide, whereas for
           electrons and
           electronic interaction the fractional loss varies with the
           PAH size. In M 82 the PAH destruction is dominated by electron 
           collisions and occurs on a very short timescale ($\sim$ thousand
           years). In Orion instead, because of the low gas temperature
           no electron and electronic contribution are possible (the rate 
           constant is close to zero, see Fig. \ref{rateConstant_fig}), and 
           hence only nuclear
           interaction is responsible for the PAH erosion, which however
           takes place on a long timescale (some 10$^{7}$ years).
           }
  \label{timeEvolution_fig}
\end{figure}

Clearly, PAHs or related larger species can survive in these
environments.  Fig. \ref{timeEvolution_fig} shows the fractional
C-atom loss, due to electron and ion (H + He + C) collisions, for two
widely different objects: the Orion Nebula (M~42) and the M~82
galaxy (cf. Table~\ref{object_tab}). The famous Orion Nebula is an
{\sc H\,ii} region with high density ($n_{\rm H}$ = 10$^{4}$
cm$^{-3}$) and low temperature ($T$ = 7000 K) gas, while M~82 is a
starburst galaxy, which shows outside the galactic plane, a
spectacular bipolar outflow of hot and tenuous gas ($n_{\rm H}$ =
0.013 cm$^{-3}$, $T$ = 5.8$\times$10$^{6}$ K).

In M~82, PAHs are completely destroyed by electrons, even for the larger PAH, 
before electronic and nuclear contributions start to be relevant. The electron 
and electronic fractional losses decrease with PAH size, thus bigger molecules 
can survive longer, while the nuclear loss is independent of $N_{\rm C}$. 
The destruction timescale is very short: after one thousand years the PAHs 
should have completely disappeared.

In Orion the situation is reversed. At the temperature considered for this 
object, electrons and electronic excitation do not contribute to PAH erosion
(cf. Fig. \ref{rateConstant_fig}). The damage is caused by nuclear 
interaction due to He collisions, with a marginal contribution from carbon
because of low abundance, and the timescale is much larger: only after 10     
million years the PAH destruction becomes relevant. Of course, we have not 
evaluated the destruction of PAHs by H-ionizing photons in the Orion {\sc H\,ii} 
region, which is expected to be very important.

The young ($\simeq 2500$ yr) supernova remnant, N132D, in the Large Magellanic 
Clouds has been studied in detail at IR, optical, UV, and X-ray wavelengths
\citep{morse95, tappe06}. A Spitzer/IRS 
spectrum of the Southern rim shows evidence for the 15-20 $\mu$m plateau -- 
often attributed to large PAHs or PAH clusters \citep{vanKer00, peeters04} -- 
and, tentatively, weak PAH emission features near 6.2, 7.7, and 11.2 $\mu$m 
\citep{tappe06}. \citet{tappe06} attribute these features to emission 
from large ($\sim 4000$ C-atom) PAHs either just swept up by the blast wave 
and not yet completely destroyed by the shock or in the radiative precursor 
of the shock. We calculate a lifetime of small ($50-200$~C-atom) PAHs in the 
relatively dense, hot gas of this young supernova remnant of $\sim 4$ months 
(Table \ref{object_tab}). In contrast, we estimate a lifetime of $150$ yr for 
4000 C-atom species (cf. Eq. \ref{timeConstant_eq} and Fig.~\ref{rateConstant1000_fig}). 
It is clear that the PAH--grain size distribution will be strongly affected in 
this environment. Given advection of fresh material into the shocked hot gas, 
the observations are in reasonable agreement with our model expectations. 
We note that the observed shift to larger PAH sizes -- so evident in the 
observations -- implies that the emitting species are not associated with 
the precursor but are actualy tracing the postshock gas and are therefore 
likely heated through collisions with the hot electrons.

This discussion clearly shows that in a low density - high temperature 
gas, small PAHs are rapidly and completely destroyed. Survival of small 
PAHs in such environnement (and so the possibility to be detected) requires a 
protective environment and/or an efficient reformation mechanism.

\section{Discussion}


\subsection{X-ray absorption}
\label{sect:x-ray}


\begin{table*}
    \caption
        {Physical properties and PAH lifetime, $\tau_{0}$, 
          in objects showing PAH emission characteristics. 
        }
        \label{object_tab}                
        \begin{center}
        \renewcommand{\footnoterule}{}      
        \begin{tabular}{c c c c c c c c c}     
          \noalign{\smallskip}
          \noalign{\smallskip}
          \hline\hline       
          \noalign{\smallskip}
          Object  &  $n_{\rm H}$  &  $T$    &  Reference   &   $\tau_{\rm 0}$(ref)  &  $\tau_{\rm 0}$(min)  &  $\tau_{\rm 0}$(max) & $\tau_{\rm object}$   &  Destruction  \\  
                  &  (cm$^{-3}$)  &   K     &             &           (yr)         &       (yr)          &        (yr)         &        (yr)          &    agent       \\
          \noalign{\smallskip}
          \hline                    
          \noalign{\smallskip}         
          M82     &   0.013     &  5.8$\times$10$^{6}$   &  (a)  &  162                  &  43     &   1302   &    $\sim$2$\times$10$^{7}$  &   electrons  \\
          M17     &   0.3       &  7$\times$10$^{6}$     &  (b)  &  8.2                  &  2      &    71    &    $\sim$1$\times$10$^{6}$  &   electrons    \\
          N132D   &   10        &  8$\times$10$^{6}$     &  (c)  &  0.3                  &  0.07   &    2.5   &    $\sim$3$\times$10$^{3}$  &   electrons         \\
          Orion   &   10$^{4}$  &  7$\times$10$^{3}$     &  (d)  &  1.3$\times$10$^{7}$   &  1.4$\times$10$^{5}$  &  7.5$\times$10$^{11}$  &  $\sim$1$\times$10$^{6}$  &  ions    \\ 
          \noalign{\smallskip}
          \hline 
          \noalign{\smallskip}
        \end{tabular}        
        \end{center}
  (a): \citet{ran08}, (b): \citet{tow03}, (c): \citet{hwang93}, (d): \citet{tielens05} \\
\end{table*}      


As our calculations show, PAHs are rapidly destroyed in the hot gas associated 
with stellar winds and supernova explosions. Any PAHs observed near such regions 
have to be isolated from this hot gas and are presumably in cold gas entrained 
in these stellar and galactic winds. However, such PAHs would still be exposed 
to energetic X-ray photons and these can be very destructive as well 
\citep{voit92, boe08}.

\begin{figure}
  \centering
  \includegraphics[width=.97\hsize]{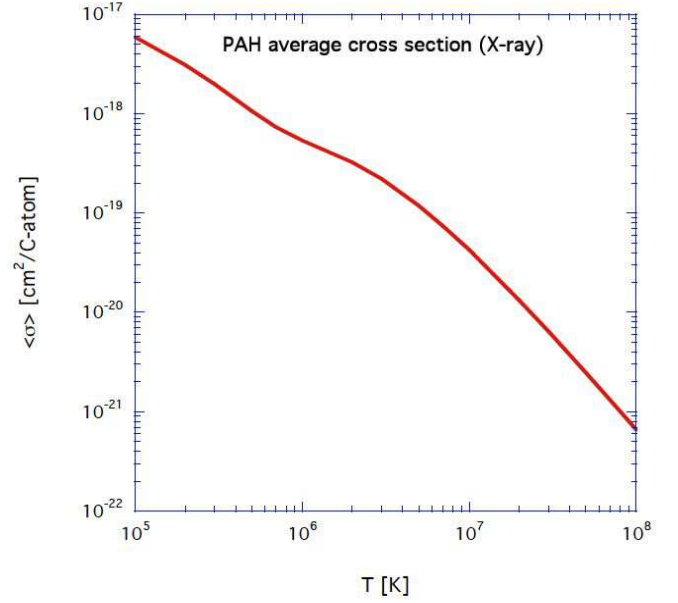}
  \caption{Planck averaged cross section of PAHs as a function of temperature. 
           The cross section is on a per C-atom basis. The strong carbon K shell 
           band edge at $\simeq 286$ eV ($2\times 10^6$ K) has been largely washed 
           out by the averaging process.
           }
  \label{Xray_fig}
\end{figure}

The photon absorption cross section of PAHs shows strong peaks at about 6 and 
17.5 eV associated with transitions involving the $\pi$ and $\sigma$ electronic 
manifolds and then another broad peak around 286 eV due to carbon K-shell 
transitions \citep{keller92, deSouza02, regier07}. Each of these peaks can show 
various subpeaks due to electronic and vibrational structure. Here, we focus on 
the high energy peak. Because we are only concerned with the Planck-averaged 
cross section, all the fine detail will be washed out and we have elected to 
evaluate the X-ray absorption rate adopting the measured absolute cross section 
of graphitic carbon taken from the NIST data base \citep{chant95, chant00}. 
Fig. \ref{Xray_fig} shows the (photon) averaged cross sections as a function 
of the black body temperature. As this figure illustrates, even the strong 
absorption edge due to the carbon K-shell is washed out by this averaging 
process, justifying our neglect of the fine detail in the absorption cross 
sections of individual PAHs. Our averaged cross section is also in good 
agreement with the recent study of \citet{boe08} for benzene.

The photon absorption rate, $R_{\rm X}$, of a PAH exposed to an X-ray photon flux, 
$N_{\rm X}$ is then given by,
\begin{equation}
  R_{\rm X}\, =\, \sigma_{\rm X}\left(T_{\rm X}\right)\, N_{\rm X}
\end{equation}
with $\sigma_{\rm X}\left(T_{\rm X}\right)$ the Planck average absorption cross section at 
temperature, $T_{\rm X}$. Consider now PAHs embedded in cold gas entrained in the 
stellar wind of the ionizing stars of the {\sc H\,ii} region M~17. The observed X-ray 
luminosity is $4\times 10^{33}$ erg s$^{-1}$ and the X-ray temperature 
is $7\times 10^6$ K \citep{tow03}. With a projected emitting area of 42 pc$^2$ 
this then implies an average X-ray photon intensity of 
$\simeq 2.6\times 10^3$ photons cm$^{-2}$ s$^{-1}$. With an average cross section 
of $\sigma_{\rm X}\, \simeq\, 9\times 10^{-20}$ cm$^2$/C-atom, we calculate a photon 
absorption rate of $7.2\times 10^{-3}$ photons/C-atom/Myr. After X-ray photon 
absorption, a PAH will typically lose a few C-atoms and, hence, the estimated 
lifetime of PAHs entrained in the stellar winds from M~17 is $\sim 50$ Myr. 
This is in excess of the stellar lifetime and the overall life time of 
the region. PAHs embedded in cold gas entrained in the galactic superwind of 
M~82 are exposed to somewhat harsher conditions. For M~82, at a distance of 
3.6~Mpc, the observed soft (0.5-2.0 keV) X-ray flux is 
$10^{-11}$ erg cm$^{-2}$ s$^{-1}$ and the temperature is 7$\times 10^6$ K. 
Evaluating the X-ray flux at a distance of 3~kpc from the nuclear starburst 
region, we calculate a photon absorption rate of $1.8\times 10^{-2}$ photons/C-atom/Myr. 
Hence, in this environment,  PAHs would be destroyed on a timescale of $\sim 20$~Myr, 
comparable to or larger than the starburst lifetime. It is clear that PAHs 
in the hot gas are predominantly destroyed by collisions with the hot gas and 
that X-ray photon absorption plays little role.

Finally, we note that here we have assumed that every X-ray photon absorption 
will lead to photodissociation. Actually, H-shell electron ejection will be 
rapidly followed by the Auger effect filling the K-shell again and simultaneously 
ejecting a second electron. This process may leave the PAH internally excited 
with typically 15-35 eV\footnote{The resulting PAH dication will quickly recombine 
with electrons and Coulomb explosion is of no consequence for such large PAHs.}.  
This energy is well above the measured appearance energy of PAH fragmentation for 
small PAHs under radiative cooling conditions (cf., Sect. 4.1) but larger PAHs may 
survive. Indeed, for an Arrhenius energy of 4.6 eV, a 50 C-atom PAH requires about 
24 eV of internal energy to dissociate with a probability of 1/2 (cf., Sect. 4.1). 
Hence, not every X-ray photon absorption will lead to fragmentation, particularly 
for large PAHs. Further experiments on larger PAHs are required to settle this issue.


\subsection{PAHs as tracers of cold entrained gas}

The results in the previous subsection illustrate that PAHs can survive in the 
surface regions of cold gas clouds embedded in a hot gas for tens to hundreds 
of million years. The expansion of a stellar or galactic wind in an inhomogeneous 
environment will in a natural way lead to the entrainment of cold gas clumps
\citep{smith84, chevalier85, hartquist93, su94, stri00, marcolini05}.
The material in these clumps will only be mildly shocked -- and PAH destruction 
by shocks is not expected as long as the shock velocity is less than about 
100~km~s$^{-1}$ (MJT) -- and can be transported with the wind while losing its 
identity only slowly, mainly through evaporation into the hot gas driven by 
thermal conduction. Multiwavelength observations of such winds support this 
global view and much of the soft X-ray and {\sc O\,vi} absorbing and emitting gas is 
now thought to originate from the interaction interfaces between the tenuous 
hot gas and the cold dense clumps \citep{chevalier85, heck02}.

A typical 0.5 keV X-ray photon will penetrate a column of about $10^{22}$ 
H-atoms/cm$^2$ and hence the `PAH-bright' surfaces of such PDRs -- which are 
typically only a column of $5\times 10^{21}$ H-atoms/cm$^2$ thick -- will be 
slowly depleted of PAHs on the timescale calculated in section~\ref{sect:x-ray}. 
However, at the same time, entrained clumps will slowly evaporate due to thermal 
conduction. Considering classical thermal conduction, the mass loss of a cloud 
due to evaporation into the hot gas is given by
\begin{equation}
  \dot{M}\, \simeq\, 1.3\times 10^{-3} \, \left(\frac{R}{\rm 10\; pc}\right)\, 
            \left(\frac{T}{\rm 6\times 10^6\; K}\right)^{5/2}
\quad {\rm M_{\odot}\; yr^{-1}}
\end{equation}
with $T$ the temperature of the hot gas and $R$ the radius of the clump \citep[cf.][]{tielens05}. 
Consider the galactic wind of M~82 and a typical structure with a size of 10 pc.  
Assuming thermal pressure equilibrium between the clump and the hot gas (eg., 
ignoring ram pressure confinement which may actually be a factor of 10 larger), 
we estimate a density of $\simeq 800$ cm$^{-3}$ for a typical PDR temperature 
of 100 K for the galactic wind region of M~82. The lifetime of such a clump is then,
\begin{equation}
  \tau_{\rm clump}\, \simeq\, 40 \, \left(\frac{n_{\rm PDR}}{10^3\,{\rm cm^{-3}}}\, \right)
                   \left(\frac{R}{\rm 10\; pc}\right)^2 \, 
                   \left(\frac{\rm 6\times 10^6\; K}{T}\right)^{5/2}\quad {\rm Myr.}
\end{equation}
Hence, `fresh' material will continuously be advected to the clump surface where 
ambient FUV photons can excite it. Thus, we conclude that such clumps can survive 
for a long time and that, proviso illumination by FUV photons, PAHs will form an 
excellent `dye' for tracing the presence of cold entrained material; or more 
specifically, for the exposed surfaces of photodissociation regions.

\subsection{Comparison with previous studies}

\citet{dwek96} determines a sputtering rate for dust in a hot dust gas 
($T \ge$10$^6$ K) that can be used to derive the lifetime, $\tau$ of a 
grain of a given radius in a hot gas of a given density, i.e.
\begin{equation}\label{dustLife_eq}
\tau \approx \frac{10^3}{n_{\rm H}} \left( \frac{a}{\rm 1 \AA} \right)  {\rm yr},
\end{equation}
where $a$ is the grain size and $n_{\rm H}$ is the gas density.
Clearly in this situation it is the smallest grain that suffer the fastest 
destruction. Hence, a treatment of the small grains, including PAHs, really 
needs to take into account the detailed physics of the interactions. For 
illustrative purposes, we can use Eq.~\ref{dustLife_eq} to show that grains
with sizes typical of PAHs ($a \approx 5$\AA) should have lifetimes, in a 
tenuous hot gas ($n_{\rm H} = 1$ cm$^{-3}$), of the order of at least a few 
thousand years. We can see that this timescale is much longer than the 
typical PAH lifetimes that we derive here, i.e. few years for a 50 carbon 
atom PAH in a gas with $n_{\rm H} = 1$ cm$^{-3}$. 
The timescale determination arrived at using Eq.~\ref{dustLife_eq} assumes 
that a PAH just behaves as a small grain. However, in the case of a grain, 
the sputtering yield is usually much less that unity because 
atoms displaced by knock-on collisions in the solid can remain in the solid. 
The very much shorter lifetimes for the PAHs that we find can be ascribed to 
the fact that the transferred energy in incident electron and ion interactions 
always leads to the loss of carbon atoms from the PAH when that energy exceeds 
the relevant binding energy. In fact the equivalent PAH `sputtering yield' can 
be greater than unity because multiple atom ejection is possible. Indeed we 
assume here that the ejected species in the case of electronic interactions are 
C$_2$ units.

The evolution of the dust and PAH thermal emission arising from a hot gas is 
affected by the changing dust size distribution. As pointed out by \citet{dwek96}, 
it is the short wavelength emission, coming from the smallest grains, that is 
most affected by dust destruction. Our work now indicates that the destruction 
timescale is much shorter than that predicted by Dwek et al. because the erosion 
of the smallest grains and PAHs is enhanced by as much as three orders of magnitude 
compared to the earlier work. Thus, the gas cooling rates derived by \citet{dwek87} 
will need to be reduced if a significant fraction of the dust mass is in the form 
of small grains and PAHs.

\subsection{C$_2$ groups loss}

The primary channel for PAH erosion in a hot gas is the progressive
loss of C$_2$H$_{n}$ ($n$ = 0, 1, 2) units following incident electron
and ion excitation of the molecule.  We note that this can only really
occur from the periphery of the PAH (see the lower part of Fig. 11 in
MJT for an illustration of this type of erosion).  In this case we can
see that a coherent aromatic structure will tend to be preserved,
which is probably not the case where the inertial sputtering of C
atoms is dominant. Thus, we conclude that PAH erosion in a hot gas
will tend to preserve the aromatic structure throughout the
destruction process.

As emphasized in MJT, PAH destruction may start with complete H-loss followed 
by isomerization to much more stable pure carbon clusters such as fullerenes 
\citep[cf.][]{tielens08}. While we have not assessed this point, we expect 
that such species will be more stable than PAHs in a hot gas.

\section{Conclusions}

We have extensively studied the stability of PAHs against electron and ion
collisions (H, He and C) in a hot gas, such as the gas behind fast non-radiative
shocks and in galactic outflows. Collisions can lead to carbon atom loss, with a 
consequent disruption and destruction of the molecule. We consider the case of a 
thermal gas with temperature $T$ in the range $10^3 - 10^8$ K.
 
An ionic collision consists of two simultaneous processes which can be treated
separately: a binary collision between the projectile ion and a single atom in the 
target (nuclear interaction) and energy loss to the electron cloud of the molecule 
(electronic interaction). 

For the nuclear interaction, as described in MJT, we have modified the existing 
theory in order to treat collisions able to transfer energy \textit{above} 
a specific threshold $T_{0}$. We adopt $T_{0}$ = 7.5 eV as a reasonable value, but 
experimental determinations of this quantity are necessary. 

The electronic energy transfer has been described in term of the stopping power of an
electron gas with appropriate electron density (jellium approximation). For 
electron collisions, we derived an analytical expression for the measured electron
stopping power in graphite and applied this to the case of PAHs. 

The dissociation probability for a PAH excited by electronic interactions and
electron collisions, has been derived using the theory of unimolecular reactions. 
The parameter $E_0$, which governs the dissociation probability, is not well 
constrained. We adopt a value of 4.6 eV consistent with extrapolations to 
interstellar conditions but better determinations, relevant to the astrophysical 
situation, are needed.

The PAH destruction process is dominated by electron collisions for gas temperatures
above $\sim$3$\times$10$^{4}$ K, and by nuclear interaction with helium below this value. 
Small PAHs are more easily destroyed than larger ones below $\sim$10$^{6}$ K, while the 
difference reduces significantly for a hotter gas. For a 1000 C-atom PAH, nuclear 
interactions are the dominant destruction process. 

In a hot and tenuous gas (e.g. M~82 galactic outflows), PAHs with sizes between 50 
and 200 C-atom are destroyed by electron collisions in few thousand years. In denser 
and colder regions (e.g. Orion), PAHs can survive for some 10$^7$ yr before being 
destroyed by nuclear interaction processes.

X-ray photon absorption plays little role in PAH destruction in the hot 
gas associated with stellar winds and supernova explosions, with respect to 
electron collisions. The PAH destruction process in a hot gas is therefore 
dominated by electron collisions. Any PAHs observed near such regions have to 
be isolated from this hot gas and are presumably in a cooler PDR-type gas 
entrained in the stellar and galactic winds. In this sense, PAHs represent a 
good tracer for the presence of entrained denser material.

Our calculated PAH lifetime in a hot tenuous gas ($T \sim$~10$^6$~K, 
$n_{\rm H} = 1$ cm$^{-3}$), is much shorter than the lifetime of an
equivalent dust grain of roughly the same size ($a \approx 5$\AA). 
Thus, might then imply that the destructive effects of ion and electron 
collisions with very small grains have previously been underestimated.
The enhanced erosion of the smallest grain and PAHs implies that the gas 
cooling rates for a hot gas ($T >$~10$^6$~K) -- which depend on the grain/PAH 
size distribution -- may need to be reduced if a significant fraction of the 
dust mass is locked in small grains and PAHs.

PAH erosion in a hot gas occurs mainly through the ejection of C$_2$ groups 
following electron collisions and electronic excitation. The C$_2$ loss occurs 
at the periphery of the molecule, thus the aromatic structure will tend to be 
preserved  throughout the destruction process.

\begin{acknowledgements}
We are grateful to L. Allamandola and L. Verstraete for useful 
discussions, and we acknowledge our referee Tom Hartquist for careful reading
and helpful comments. E.R.M. thanks G. Lavaux for support and technical 
assistance and acknowledges financial support by the EARA 
Training Network (EU grant MEST-CT-2004-504604).
\end{acknowledgements}


\end{document}